\documentclass[prl,twocolumn,superscriptaddress,floatfix,nofootinbib]{revtex4-1}
\usepackage{graphicx}
\usepackage{color}
\usepackage{epsfig}
\usepackage[caption=false]{subfig}

\begin{document}

\title{Effect of pinning on the yielding transition of amorphous solids}
\author{Bhanu Prasad Bhowmik}
\affiliation{Tata Institute of Fundamental Research, 
36/P, Gopanpally Village, Serilingampally Mandal,Ranga Reddy District, 
Hyderabad, 500107, Telangana, India}
\author{Pinaki Chaudhuri}
\affiliation{Institute of Mathematical Sciences, IV Cross Road, CIT Campus, Taramani,
Chennai, 600113, Tamil Nadu, India.}
\author{Smarajit Karmakar}
\affiliation{Tata Institute of Fundamental Research, 
36/P, Gopanpally Village, Serilingampally Mandal,Ranga Reddy District, 
Hyderabad, 500107, Telangana, India}
\begin{abstract}
Using numerical simulations, we have studied the yielding response, in the athermal quasi static limit, 
of a model amorphous  material having inclusions in the form of randomly pinned particles.
We show that, with increasing pinning concentration, the plastic activity becomes more spatially  
localized, resulting in smaller stress drops, and corresponding increase in the magnitude of strain
where yielding occurs. We demonstrate that, unlike the spatially heterogeneous and avalanche led
yielding in the case of the unpinned glass, for the case of large pinning concentration, yielding
takes place via a spatially homogeneous proliferation of localized events. 
\end{abstract}
\maketitle

{\em Introduction}:  Amorphous solids are ubiquitous in nature and their mechanical properties are useful in 
diverse practical applications \cite{barrat_lemaitre,rodneyreview2011,bonnreview}. Such solids yield beyond  a stress or 
strain threshold, and this yielding behaviour is fundamental to many such applications. However, systematic understanding of
the microscopic processes leading to yield is still missing. 

It has been demonstrated that the elementary events leading to yield correspond to local plastic activity within
a {\em shear transformation zone}, wherein a small set of particles undergo irreversible structural rearrangement  \cite{stz}. 
Starting from an initially quiescent glassy state, 
shear initially induces few such plastic events, which with increasing strain, proliferate and after some strain,
the system fails to resist applied shear and starts to yield and rupture (brittle) or flow (ductile). Very recently,
it has been demonstrated that yielding via a brittle or ductile response can be achieved by the degree
of annealing undergone during glass formation \cite{ozawa2018}, and the ductile rupture corresponds
to a first-order nonequilibrium transition.

Most theoretical studies probing the yielding transition have worked in the athermal quasi-static limit, which 
mimics the deformation of the system at zero temperature and strain rate
$\dot\gamma \to 0$ \cite{barrat_lemaitre}. It has been evidenced
that the stress released during plastic activity results in a cascade of events leading to 
catastrophic system  spanning  avalanches \cite{barrat_tanguy, lemaitre_maloney}.
Focusing on steady-state behaviour, following yielding from the quiescent state,  one
observes power law dependence of energy drop ($\Delta U$) and stress 
drop ($\Delta \sigma$) across the plastic rearrangements in the steady 
flowing state as $\Delta U \sim N^{\alpha} \quad \mbox{and} \quad \Delta \sigma \sim N^{\beta}$,
where these exponents are found to have universal value $\alpha = 1/3$ 
and $\beta = -2/3$, irrespective of model and spatial dimensions \cite{karmakarpre2010}. 
Scale free nature of these avalanches indicates some type of criticality in the yielding process,
and such exploration of critical behaviour has also been extended to the regime of
finite, but small, shear-rates \cite{lemaitre2009, salerno, lin, liumartens2016}
and finite temperatures \cite{gps2016}.
Further, it has been proposed recently \cite{PPRSPNAS2017} that the nature of this critical 
behavior is manifested as a spinodal point of an underlying thermodynamic 
phase transition, described by an approproate replica ``order parameter".

In this Letter, we probe the quasistatic elastoplastic behaviour of the amorphous solid, altered by the presence of 
tiny inclusions, in the form of pinned particles. Recently, such random pinning
has been found to be an interesting tool to test different theories of glass
transition and for probing the growth of static 
structural order in the system \cite{cammarota2012, kob2013, ozawa2015, sourish2015a, sourish2015b, raj2017}.
For our present study of the shear response, we consider the case where the tiny inclusions undergo affine deformation, 
when the macroscopic solid is deformed, but do not have any non-affine motion. 
While it is historically known that such inclusions strengthen a material \cite{torquato_book, hofman2008, ratul2013, gendelman2014}, 
a systematic statistical study of yielding and its microscopic ramifications is still missing, except for 
some  investigations via mesoscale or continuum models \cite{homer2015, tyukodi2016}.

Our work shows that as the solid becomes more and more rigid, 
with increase in the concentration of pinned particles, 
the initiation of plastic activity gets delayed too,
and thereby the strain at which yielding occurs systematically shifts to higher values.   
The drop in stresses, corresponding to plastic activity, also become smaller in scale, with increased concentration, 
and this is related to  relaxation processes becoming more localized. Consequently,
the stress statistics reveal that drop sizes change from sub-extensive to 
intensive in system size, with increasing pinning concentration. 
Therefore, in contrast to largescale spatially heterogeneous avalanches, 
as has been observed in usual amorphous system, 
for pinned systems, the yielding occurs via the homogeneous
accumulation of localized plastic activity.

{\em Model and method}: For our study, we consider the
well known glass forming liquid, the 
Kob-Andersen Model in both two and three dimensions \cite{bruning2009, SI}. 
The  glassy states, whose mechanical response we study,
are prepared by first equilibrating the system at high temperature ($T = 1.0$)
and then cooling it to a lower temperature $T = 0.01$ using a cooling rate 
$0.01$ per MD steps and  then further quenched to the inherent structure using 
conjugate gradient (CG) minimization \cite{SI}. The pinned states are generared by
freezing a fraction of particles within the states  reached at $T = 0.01$.
In our study, the fraction of randomly pinned particles  varies in the range $c\in [0, 0.15]$. 

For athermal quasistatic (AQS) simulation, we have started from energy
minimum state and then strained the system with incremental strain 
$\delta\gamma = 1\times10^{-5}$. At each step the system is minimized using 
CG minimization. While calculating the energy drop and stress drop statistics,
we estimated the precise location of the plastic drop using incremental strain 
$\delta\gamma = 1\times10^{-6}$ \cite{SI}. 
During the shearing process, the randomly pinned particles are allowed to
deform similar to other particles when the system is being overall deformed by an affine 
transformation, but, during minimization which causes non-affine displacements,
pinned particles remain frozen in space.


\begin{figure}
\begin{center}
\includegraphics[width=0.47\textwidth]{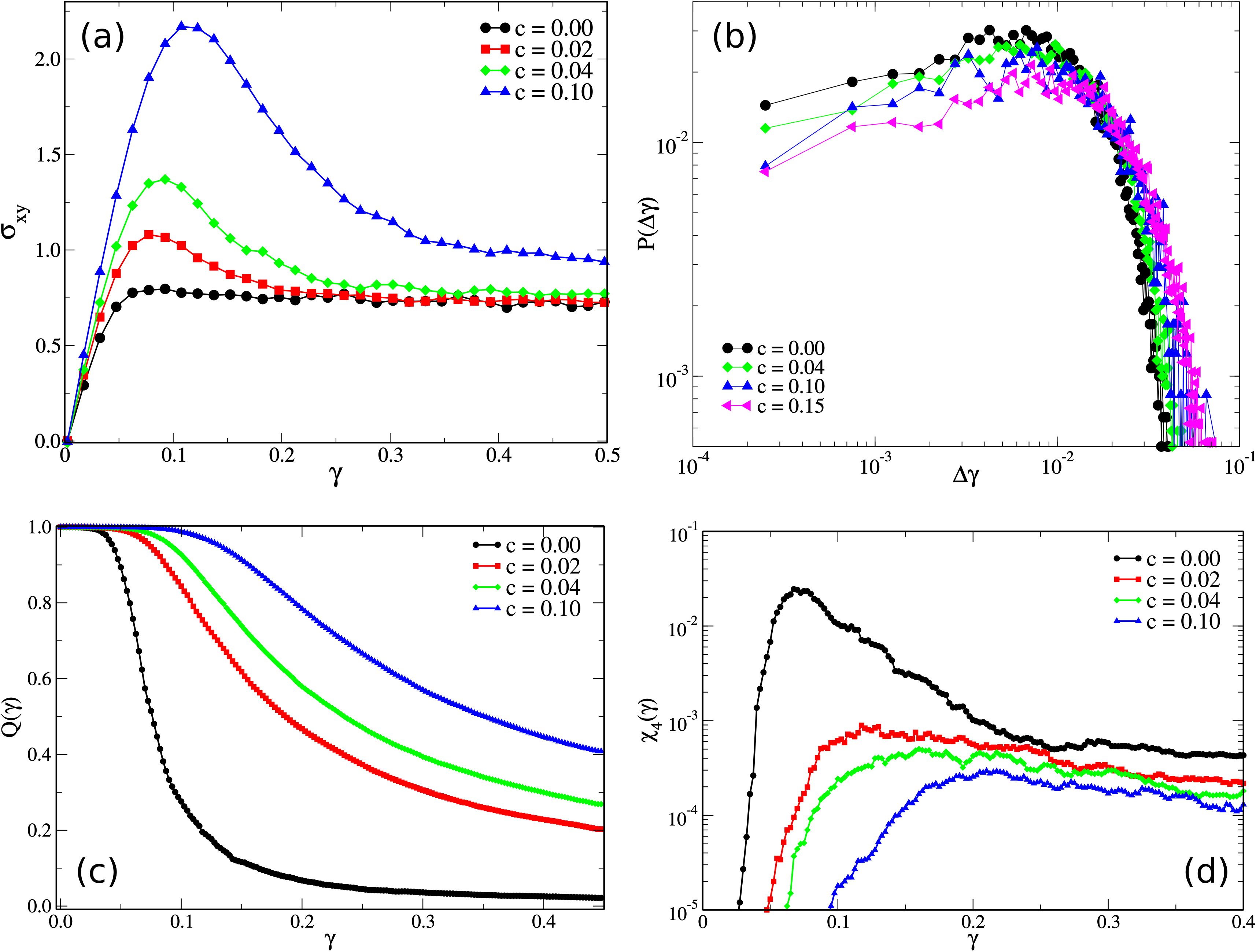}\\
\caption{{\bf  Macroscopic monitoring of yielding.} (a) Evolution of shear stress, $\sigma_{xy}$ with increasing strain $\gamma$, for
different pinning concentrations $c$. (b)  Distribution of 
$\Delta\gamma$, the strain interval between two successive stress drops, during transient response from quiescent
state.  (c) Corresponding evolution of overlap function $Q(\gamma)$ and (d) related fluctuations $\chi_4(\gamma)$, 
computed over an ensemble
of initial states, the peak location of which corresponds to the yield point ($\gamma_Y$). 
}
\label{fig1}
\end{center}
\end{figure}

{\em Macroscopic scenario of shear-induced yielding of a quiescent glass}: 
We first illustrate the macroscopic mechanical response of the systems by
considering the evolution of the shear stress as a function of the increasing
applied strain. The corresponding data is plotted in Fig.\ref{fig1}(a), for $N = 4000$ particles,
and for different concentration of pinning ($c$), with the averaging being done
for an ensemble of independent states, for each value of $c$. 
Initially, the stress increases linearly
with strain, followed by an intermediate regime of nonlinear response, before
eventually large scale plasticity sets in and the system reaches long-time steady flow.
For the case of unpinned glass, there is not much of a stress overshoot, prior to
the onset of steady flow, as is the usual case in most moderately annealed glasses \cite{ozawa2018}.
However, as the pinning concentration is increased, a stress overshoot appears
and the stress peak increases with increasing $c$, with the location of
the peak occuring, also, at larger values of strain. 
Meanwhile, at small strain, the slope of the stress versus strain curve,  also, becomes steeper,  
indicating that the pinning is making  the solid more stable  leading to increase in the shear modulus, as reported earlier \cite{ratul2013}. 

This increased rigidity with pinning is also reflected in  the distribution of 
the strain interval between two successive initial stress drops, $\Delta \gamma$,
as shear is applied to the quiescent glassy states \cite{karmakarpre2010}. 
As shown in Fig.\ref{fig1}(b), for different pinning concentrations, the probability of having the  
first plastic drop at small strain decreases, with increasing $c$, and more weight 
is transferred to larger strain intervals, 
demonstrating that the system becomes more stable with increasing pinning concentration. 

To quantitavely identify the value of the yield strain as a function of the pinning concentration, we compute the fluctuation, 
$\chi_4 (\gamma)$, of the overlap function $Q(\gamma)$, over the ensemble of
initial states, for increasing values of $c$; 
see Supplementary Material \cite{SI} for further details. 
$Q(\gamma)$ is computed,  only in reference to each initial quiescent state within
the ensemble for a particular $c$.
The average $Q(\gamma)$ computed
over the independent trajectoires starting from these initial states, as shown in Fig.\ref{fig1}(c), shows that the structural relaxation slows
down with increasing pinning concentration. The 
corresponding evolution of $\chi_4(\gamma)$
as a function of increasing strain (see Fig.\ref{fig1}(d)) has a non-monotonic behaviour
for all values of $c$. The location of the peak is identified as the strain threshold at which the system, for each pinning concentration, yields \cite{ozawa2018}.  
The data shown in  Fig.\ref{fig1}(d) quantitatively
demonstrates that the strain value at yielding increases with increasing pinning concentration. 
We also note here that the scale of fluctuations (i.e. peak height of $\chi_4 (\gamma)$) decreases
with $c$, as has also been observed in equilibrium dynamics \cite{kob2014}, reflecting that pinning exerts constraints on
the possible explorable states, for the system.


\begin{figure*}
\begin{center}
\includegraphics[width=0.90\textwidth]{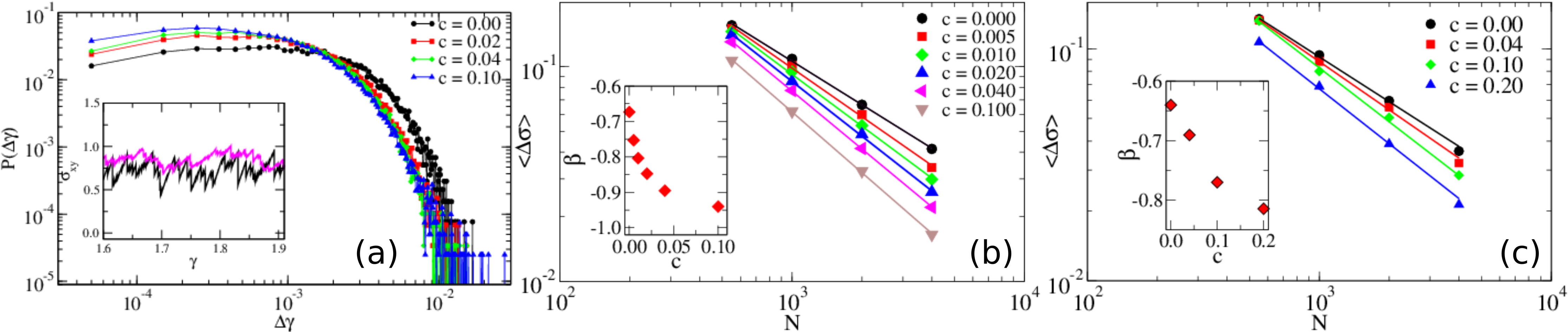}\\
\includegraphics[width=0.90\textwidth]{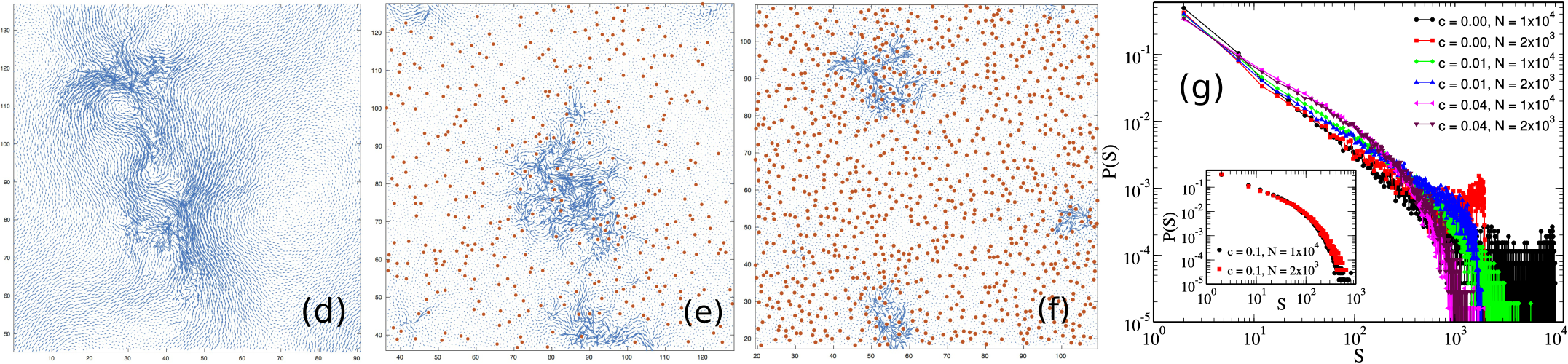}\\
\caption{{\bf Steady state.} ({\sf Top}) {\em Macro behaviour.} from left to right --
(a) Distribution of 
$\Delta \gamma$, the strain interval between two successive stress drops during
steady state. The inset shows the stress vs strain curves for the unpinned (c=0)
and pinned (c=0.15) cases. (b)-(c) Finite size effects in stress drops
($\Delta \sigma \sim N^{\beta}$), for d=2 (ii) and d=3 (iii),
with each inset showing the corresponding evolution of exponent $\beta$ with
pinning concentration $c$. 
({\sf Bottom}) {\em Microscopic scenario.} from left to right -- (d)-(f) Displacement field, during
a single large stress drop, with changing pinning concentration. (g) Distribution of cluster size, $P(S)$, of active particles for various pinning concentrations,
for different system sizes.}
\label{fig2}
\end{center}
\end{figure*}


{\em Macroscopic steady state response}: We now focus on the steady-state quasi-static flow response of the system, in the
absence and presence of pinning. Typical time-series of the stress, in steady state, for single trajectory, is shown in the
inset of Fig.\ref{fig2}(a), for $c=0$ and finite $c$: for the unpinned case, the stress drops have the well-known saw-tooth profile \cite{barrat_lemaitre},
and for increasing pinning, the stress drops become visibly smaller.  If we consider the probability distribution of $\Delta\gamma$, the  strain 
gap between two consecutive plastic drops in the steady state, we observe [see Fig.\ref{fig2}(a)] that 
the probability of having plastic events within small $\Delta\gamma$ in the steady state increases with increasing pinning  concentration,
whereas the probabity of larger $\Delta\gamma$ between stress drops decreases. This suggests that, in the steady state, the system has  strong tendency 
to have many small plastic drops instead of large system spanning avalanches.

We have analyzed the statistics of drops in shear stress, $\langle{\Delta\sigma}\rangle$, during steady state, as a
function of system size for different pinning concentration. 
In Fig.\ref{fig2}(b)-(c), we probe $\Delta \sigma \sim N^{\beta}$
for two and three dimensional systems, respectively. It is known that, for  different models for the unpinned system,
the exponent $\beta$ is universal, viz. $\beta = -2/3$ \cite{karmakarpre2010}. In our study, as seen in the insets of Fig.\ref{fig2}(c)-(d),
we find that  $\beta$ is  a strong function of pinning concentration. With increasing pinning 
concentration, $\beta$ goes from $-2/3$ to close to $-1.0$ for two dimensional 
model with $20\%$ change in pinning concentration. The same in three dimension 
changes somewhat weakly, going from $-2/3$ to somewhat below $-0.8$, within
the same amount of increase in pinning concentration. This is to some extent 
expected as there will be more paths available in three dimensions for 
avalanches to percolate the whole system.  In any case, here, we can conclude
that, with increasing pinning concentration, stress drops are changing from the sub-extensive
nature to being intensive in system size. This 
leads one to guess that the avalanche are probably being affected by the random pinning.

\begin{figure*}
\begin{center}
\centerline{\includegraphics[scale=0.4]{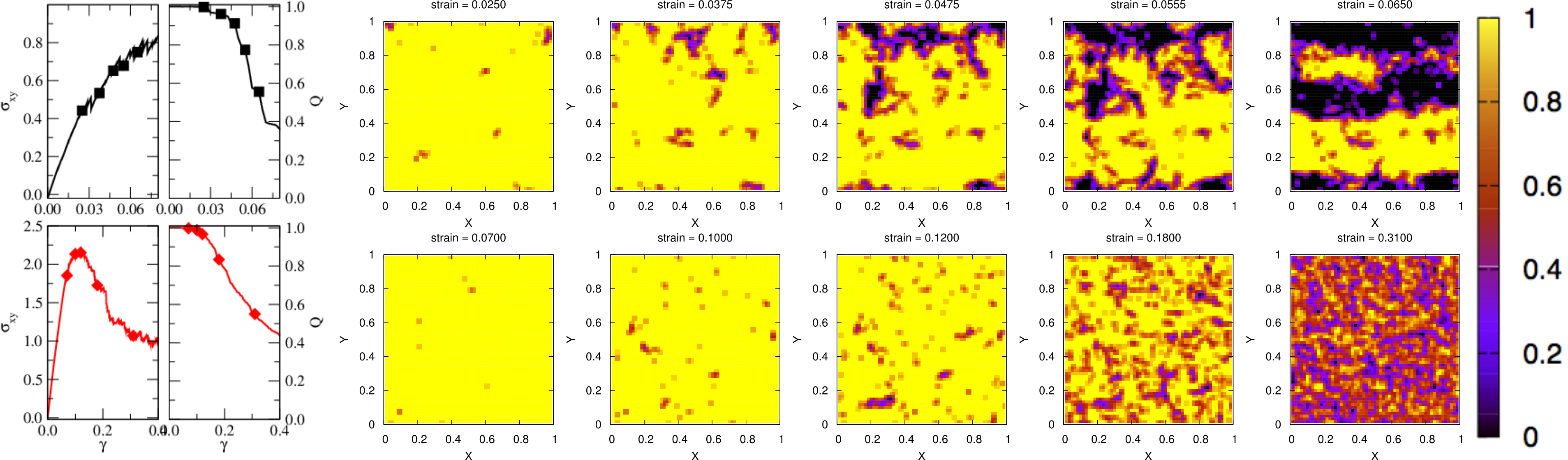}}
\caption{{\bf Microscopics of yielding.} {$N=10000$. Sequence of maps of local overlap, $\tilde{Q}(x,y)$, during
yielding, for (top) $c=0$, shown for strain values of 0.0250, 0.0375, 0.0475, 0.0555, 0.0650, from left to right, (bottom) $c=0.10$,
shown for strain values of 0.0700, 0.1000, 0.1200, 0.1800, 0.3100, from left to right.  
Corresponding evolution of macroscopic stress ($\sigma_{xy}$) and also global overlap function ($Q$),
with increasing strain, are also show in the left end of top and bottom panels. The colorbar shows the scale of local overlap, for each case.}}
\label{fig3}
\end{center}
\end{figure*}

{\em Microscopic steady state response}: After having characterized the macroscopic response and its
variation with changing pinning concentration, we now delve into analyzing how the mechanical response
to the applied shear manifests at the microscopic scale. In order to do that, we study the 
displacement field generated during stress drops, and in particular, we consider typical cases where the large
stress drops are observed, for both pinned and unpinned scenarios.  In Fig.\ref{fig2}(d)-(f), we show such displacement fields, 
generated during steady shear, for various scales of pinning. For the unpinned case (Fig.\ref{fig2}(d)), a large stress drops 
corresponds to large scale displacements, in the form of an avalanche, spanning the entire system, as has been observed and well-studied 
earlier \cite{barrat_tanguy, lemaitre_maloney, barrat_lemaitre, rodneyreview2011}.  However, as we increase
the pinning concentration, the spatial scale of the region, undergoing large displacements, shrinks (see Fig. \ref{fig2}(e) 
for $c=0.05$) and eventually becomes completely localized for larger values of $c$ (see Fig. \ref{fig2}(f) for $c=0.10$). 
Thus, the propagation of stress, following the plastic reorganization, gets hindered due to the presence of the pinning centres. 
Therefore, the avalanche-like character of
the non-affine displacements, during a stress drop, is progressively cut-off, with increasing pinning concentration.

Further, we now quantify this observation, by identifying
the particles that are part of the relaxation process during the stress drop
and measuring the cluster size corresponding
to these mobile particles. In order to label a particle as mobile,
we only consider those that move more than some threshold distance ($r_c = 0.05$),
during the drop in stress, by analyzing the 
probability distribution of displacements of particles during plastic 
events show power law behaviour with an exponential tail \cite{SI}. 
We, then, study the 
nature of  the clusters formed by these particles during a plastic event. 
In Fig.\ref{fig2}(h), we plot distribution of cluster size of  mobile 
particles for two different  system size $N = 2000$ and $N = 10000$. The
distribution for $c=0$ shows a power law behavior with a hump 
at large cluster size. This is the sign of presence of percolating cluster,
which is visible  for both the system sizes. Note that for $N = 10000$, the hump 
appears at a cluster size, $S$, that is larger than that for the $N = 2000$. This is completely consistent with 
the expectation that avalanches grow with  system size as $\sim N^{\alpha}$. 
For higher pinning concentration  $c = 0.10$, the range of power law decreases and shows no hump at large 
cluster size. Also, the distributions for two different system size do 
not show any size effect. 
This evidences the occurrence of localized plastic 
events for higher pinning concentration. 

Thus, via all these statistical analyses, we can conclude that system spanning avalanches start to  
becomes less frequent with increasing pinning concentration and beyond  a certain pinning concentration, 
the avalanche like cascade of plastic rearrangements completely disappear.

{\em Microscopic scenario for yielding}: 
Having had a microscopic idea of the nature of particle displacements during a stress drop,
we now return to the question of how such stress drops, and consequently particle displacements, lead to yielding
and eventual largescale flow.  To monitor this, we construct maps of local overlap, $\tilde{Q}(x,y)$, \cite{SI}, relative
to the initial queiscent state, and follow how the map evolves as a function of increasing strain, for the pinned and unpinned
case. This evolution is shown in Fig.\ref{fig3}, for a trajectory from either 
case, for a system of $N=10000$. Additionally, we show the corresponding evolution of shear stress ($\sigma$) and overlap ($Q$) with strain. 

For the unpinned case,  shown in the top panel of  Fig.\ref{fig3}, at early strains, we have the first plastic events, occurring at small scale, 
at different spatial locations. As strain increases, more such events occur, 
with some of the relaxed regions 
(which, now, have lower local overlap relative to the initial state) increasing in scale, i.e. they become nuclei of yielding in their neighbourhood. 
Eventually these regions connect, avalanches occur and  the relaxed parts span the system, i.e. there is a percolation of mobile regions \cite{gps2016}. 
Importantly, one sees the formation of shear-band like structures, spanning the system in the shear-direction, and the spatial scale of these bands widen with increasing strain. 
To summarize,  in this case, the states, transient to complete fluidization, correspond to avalanches and flow heterogeneities. 

We now demonstrate the contrasting scenario for the case of pinning and, 
for that, we consider,  the case of $c=0.1$, where
we have shown that stress drops lead to localized plasticity and not avalanches. 
The corresponding maps of local overlap, are shown in the bottom panel of Fig.\ref{fig3}. In 
this case, the first visible signs of plastic activity occur at larger strain,
since the pinning has rendered the system more rigid. With increasing strain, 
these regions start to grow in number.
This is in contrast to the unpinned case, where the initial regions were driving increased plasticity, in their neighbourhood. In the case of pinned particles,
since plasticity is localized, non-local effects are screened, which lead to more regions locally yielding. Because of such a
scenario, these spots pop up in a spatially homogeneous manner,  and no system-spanning events are observed. 
Thus, at larger strain, yielding happens via the homogeneous occurrence of such small-scale localized activities, filling up the 
entire system. We note the contrasting spatial picture for the two systems, at the right-most end of the
two panels, which both have the same global overlap value ($\approx 0.57$). Further, we also note here, that even though there 
is a prominent stress overshoot, for $c=0.1$, and that too in the vanishingly small shear-rate
limit, no flow heterogeneities are observed in this case, as has often been conceived \cite{moorcroft2011}.

To summarize, we have studied the shear response of amorphous solids,
in the quasistatic limit, with point-like inclusions embedded within
them. The significant finding is that, with increasing concentration,
yielding gets delayed as the initiation of plastic activity becomes
difficult with the local constraints imposed by the pinned particles. One
would assume that the local yielding thresholds \cite{patinet2016}
get altered with the presence of these inclusions, 
leading to this hindrance.  Further, even when
plastic events occur, the propagation of the stress relaxation across the
system is cut off by the inclusions, leading to localized dissipation
in the form of smaller stress drops, which is very different from the
avalanche like dissipation in the unpinned amorphous solid. From the
generated displacement fields, it is clear that the stress propagator
which has the quadrupolar Eshelby form in the case of unpinned solid,
becomes modified as the pinning concentration increases and the exact
form perhaps needs to be calculated, to develop appropriate mesoscale
description \cite{elastoreview2017}
for the ensuing yielding scenario.  Finally, we have brought together these different
elastoplastic aspects to illustrate how yielding and subsequent
plasticity proceeds, starting from a quiescent state. Distinct from
the spatially heterogeneous initiation of largescale plasticity in the
unpinned amorphous solid, the presence of the inclusions leads to a more
spatially homogeneous yielding, with the different zones of activity
likely to be more decoupled and toppling independently,
as evidenced by the statistics of stress drops that emerge from 
our investigation. Thus, even though there is increased
rigidity via the inclusions, there is no brittle-like rupture with
accompanying shearbands as has been observed for ultrastable 
glasses where slower annealing leads to increased rigidity \cite{ozawa2018}. 
How this scenario
influences the rheology at finite shear-rates \cite{lin,liumartens2016,
chaudhuri2012} remains to be investigated.

{\em Acknowledgements}: We would like to thank Surajit Sengupta, Juergen Horbach, Jean-Louis Barrat, Kirsten Martens and Srikanth Sastry
for useful discussion.


\end{document}


\title{ Role of Avalanches in the Yielding Transition of Amorphous Solids}
\author{Bhanu Prasad Bhowmik$^{1}$}
\author{Pinaki Chanudhuri$^{2}$}
\author{Smarajit Karmakar$^{1}$}
\email{smarajit@tifrh.res.in}
\affiliation{
$^1$ Tata Institute of Fundamental Research, 
36/P, Gopanpally Village, Serilingampally Mandal,Ranga Reddy District, 
Hyderabad, 500107, Telangana, India \\
$^2$ Institute of Mathematical Sciences, IV Cross Road, CIT Campus, Taramani,
Chennai, 600113, Tamil Nadu, India.}
\maketitle
\noindent{\bf Models and Simulation Details:}
We have preformed simulation of well-known glass forming liquid, the 
Kob-Andersen Model in both two and three dimensions \cite{bruning2009}. It is a binary 
mixture with 80\% large particle (A) and 20\% smaller particles (B) 
in three dimension. In two dimension the ratio of A and B particles 
is slightly different, A:B = 65:35. The particles interact via LJ 
potential given below 
\begin{equation}
V_{\alpha,\beta} = 4\epsilon_{\alpha\beta} [(\sigma_{\alpha\beta}/r)^{12} - (\sigma_{\alpha\beta}/r)^6],
\end{equation}
where $\alpha$ and $\beta$ stand for two different types of particle A and B. 
$\epsilon_{AA}$ = 1.0, $\epsilon_{AB}$ = 1.50, $\epsilon_{BB}$ = 0.50. 
$\sigma_{AA}$ = 1.0, $\sigma_{AB}$ = 0.88, $\sigma_{BB}$ = 0.80.  

\vskip +0.2cm
\noindent{\bf Preparation of Sample:}
For the preparation of initial states, first the system is 
equilibrated at high temperature $T = 1.0$, 
then  brought down to a lower temperature $T = 0.01$ using a 
cooling rate $10^{-4}$ per MD step and then quenched to the 
inherent structure using conjugate gradient (CG) minimization.
\vskip +0.2cm
\noindent{\bf Shear Protocol:}
For athermal quasistatic (AQS) simulation, we have started from a 
inherent state and then strained the system with incremental strain 
rate $\delta\gamma = 5\times10^{-5}$. At each step the system is 
minimized using CG minimization. While calculating the energy drop 
and stress drop statistics, we estimated the precise location of 
the plastic drop using incremental strain rate 
$\delta\gamma = 2\times10^{-6}$. Thus the uncertainty in our 
estimate of strain interval between two successive plastic drop 
is of the order of $\delta\gamma \sim 2\times10^{-6}$. We also believe 
that the uncertainty in the estimate of energy and stress drops are 
also of the order of $\delta U \sim 5\times10^{-5}$ as shear 
modulus in these model systems are $\mu \sim 10$. 

\vskip +0.2cm
\noindent{\bf Pinning Protocol:}
A fraction of particles are chosen randomly from the inherent 
configurations. During the shearing process, the randomly pinned 
particles are allowed to deform as other particles while the 
system is being deformed by affine transformation but during 
minimization which causes non-affine displacements pinned particles 
remain frozen at their position. In our study the fraction of 
random particles changes in the range $c\in [0, 0.15]$ for 2d 
and $c\in [0, 0.20]$ for 3d. As the particles do not move during 
minimization it is possible that because of affine transformation 
they may come very close to each and energy will become very large. 
To avoid this situation the randomly pinned particles are chosen 
such a way that the following condition is satisfied: 
$y_i-y_j \ge r_c$, where $y_i$ is the y component of the position 
vector of $i^{th}$ particle. $r_c$ is the distance where repulsive 
part of interaction becomes zero.
\vskip +0.2cm
\noindent{\bf Overlap Function:}
We have computed the overlap function $Q(\gamma)$ studied in 
a recent work \cite{stz}. The overlap function gives the 
information of about the similarity of two different configurations 
that are separated by strain $\gamma$. The self part of 
$Q(\gamma)$ can be written as 
\begin{equation}
Q(\gamma) = \frac{1}{N(1-c)}\sum_{i=1}^{N(1-c)}w(\left| r_i(0) - r_i^{NA}(\gamma)\right|),
\end{equation}	
where the window function $w(x) = 1$ when $x\leq a$ and $0$ otherwise. 
We choose $a = 0.3$. $\{\vec r_{i}(0)\}$ is the configuration at zero 
strain and $\{\vec{r}_{i}^{NA}(\gamma)\}$ is non affine configuration 
at a particular strain $\gamma$. $N$ is the total number of particle 
in the system and c is pinning concentration. We have done simulation 
for $N = 4000$ to compute  $Q(\gamma)$. $\{\vec{r}_{i}^{NA}(\gamma)\}$ 
is obtained by adding only the non-affine displacements of the 
particles and removing the affine displacements after each incremental 
change in strain via affine transformation. The fluctuation of overlap 
function $\chi_{4}(\gamma)$ can be computed using the following 
expression 
\begin{equation}
\chi_4(\gamma) = \frac{1}{N(1-c)}
[\langle Q^2(\gamma)\rangle] - \langle Q(\gamma) \rangle^2] \,
\end{equation}
where $\langle\ldots\rangle$ represents averaging over many 
statistically independent simulation runs.

\begin{figure}[h]
\begin{center}
\includegraphics[width=0.5\textwidth]{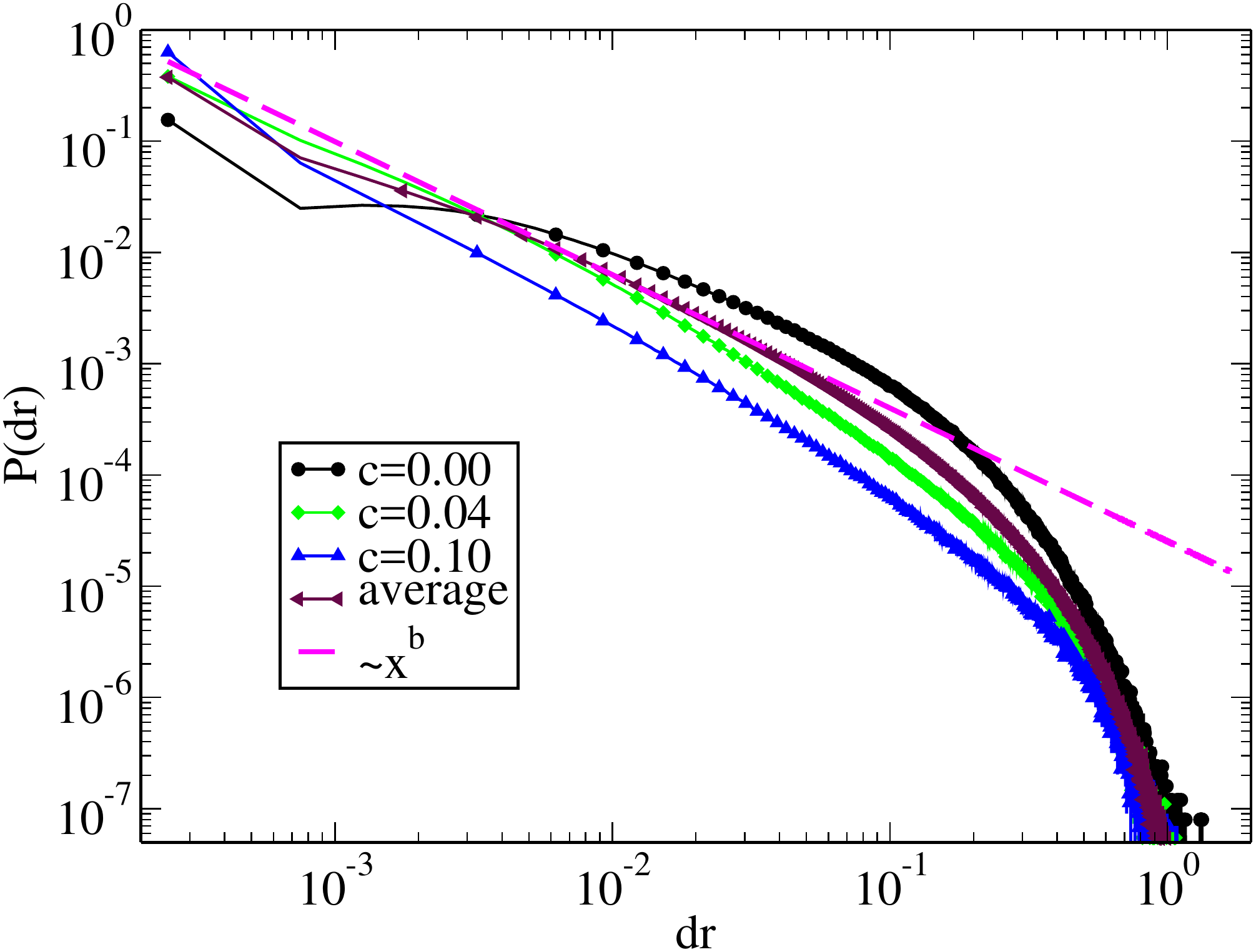}
\caption{Distribution of displacement of particles across plastic 
drops for different pinning concentration. Notice the initial 
power law behaviour that signifies the elastic deformation of 
the particles. The value at which the distribution deviates 
from power law behaviour, is defined as a cut off distance for 
identifying the plastically deformed particles.}
\label{cutoff}
\end{center}
\end{figure}

\vskip +0.2cm
\noindent{\bf Local Overlap Function:}
To study the spatial variation of overlap function we have computed 
$\tilde{Q}(x,y,\gamma)$, using $N = 10000$. We divide the system into $60\times60$ grids and 
computed the overlap function for each cell and then observe how 
this local overlap function evolves as a function of strain. The self 
part of overlap function for $i^{th}$ cell is given by 
\begin{equation}
\tilde{Q_{i}}{(x,y,\gamma)}= \frac{1}{(n_{i} - c_{i})}\sum_{i=1}^{n_{i}-c_{i}}w(\left| r_i(0) - r_i^{NA}(\gamma)\right|).
\end{equation}  		
$n_{i}$ and $c_{i}$ are the number of unpinned particle and pinned 
particles present inside the $i^{th}$ cell respectively when strain 
is zero. $\{\vec r_{i}(0)\}$ represents the configuration of particles 
present in the $i^{th}$ cell for $\gamma = 0$ and 
$\{\vec{r}_{i}^{NA}(\gamma)\}$ represents the non-affine configuration 
of the particles which were inside the same cell at unstrained state 
for a particular strain $\gamma$. While dividing the simulation box 
into cells we made sure that each cell should contain at 
least one particle.  
\vskip +0.2cm
\noindent{\bf Identification of plastically deformed particles:}
We follow the following procedure to identify the particles that 
under go plastic deformation during shear deformation. If a particle
moves more than a critical cut off distance, $r_c$ when the system 
is subjected to plastic deformation, we define that particle to be
plastically deformed or active particle. To calculate $r_{c}$ we first
look at the distribution of displacement of all the particles for all 
pinning concentration averaged over many plastic drops for small strain. 
System with higher pinning concentration has been strained up to higher 
external strain to get comparable number of plastic drops as yield point 
shifts to larger strain for these systems. The distribution shows power 
law dependence as shown in Fig.\ref{cutoff} with a sharp exponential 
drop due to finite size effects. We also find that the power law regime
survives up to larger displacement with increasing pinning concentration.
This suggests that presence of pinning makes system more elastic. To get 
the absolute value of $r_{c}$ we take the average distribution shown in 
Fig.\ref{cutoff} and fit the power law part of the distribution as
$f(x) \sim x^b$ within a range $[0.005,0.05]$. For displacements 
larger than $0.05$, the distribution deviates from power law and we 
choose $r_{c} = 0.05$. A different choice of cutoff does not change the 
results qualitatively.
